\documentclass[twocolumn,showpacs,preprintnumbers,amsmath,amssymb,floatfix]{revtex4}

\usepackage{graphicx}
\usepackage{dcolumn}
\usepackage{bm}

\begin{document}

\preprint{APS/123-QED}

\title{Turbulent kinetic energy in the energy balance of a solar flare}

\author{E. P. Kontar$^{1}$, J. E. Perez$^{1,5}$, L. K. Harra$^{2}$, A. A. Kuznetsov$^{3}$,  A. G. Emslie$^{4}$, N. L. S. Jeffrey$^{1}$, N. H. Bian$^{1}$, \&  B. R. Dennis$^{6}$}
\affiliation{$^{1}$School of Physics \& Astronomy, University of Glasgow, Glasgow G12 8QQ, UK}
\affiliation{$^{2}$UCL-Mullard Space Science Laboratory, Holmbury St Mary, Dorking, Surrey, RH5 6NT, UK}
\affiliation{$^{3}$Institute of Solar-Terrestrial Physics, Irkutsk, Russia}
\affiliation{$^{4}$Department of Physics \& Astronomy, Western Kentucky University, Bowling Green, KY 42101, USA}
\affiliation{$^{5}$ Faculty of Physics and Mathematics,
Autonomous University of Nuevo Leon, San Nicolas de Los Garza, Mexico}
\affiliation{$^{6}$ Solar Physics Laboratory, NASA Goddard Space Flight Center, Greenbelt, MD 20771, USA}
\date{\today}

\begin{abstract}
The energy released in solar flares derives from a reconfiguration of magnetic fields to a lower energy state, and is manifested in several forms, including bulk kinetic energy of the coronal mass ejection, acceleration of electrons and ions, and enhanced thermal energy that is ultimately radiated away across the electromagnetic spectrum from optical to X-rays. Using an unprecedented set of coordinated observations, from a suite of instruments, we here report on a hitherto largely overlooked energy component -- the kinetic energy associated with small-scale turbulent mass motions. We show that the spatial location of, and timing of the peak in, turbulent kinetic energy together provide persuasive evidence that turbulent energy may play a key role in the transfer of energy in solar flares.  Although the kinetic energy of turbulent motions accounts, at any given time, for only $\sim (0.5-1)$\% of the energy released, its relatively rapid ($\sim$$1-10$~s) energization and dissipation causes the associated throughput of energy (i.e., power) to rival that of major components of the released energy in solar flares, and thus presumably in other astrophysical acceleration sites.
\end{abstract}

 \maketitle


During a solar flare, up to $10^{32}$~ergs of energy stored in magnetic fields in the the solar corona is converted into the energy of accelerated particles,
bulk flows, and heating \cite{1988psf..book.....T}.
Observations \cite{2008LRSP....5....1B,2011SSRv..159..107H,2012ApJ...759...71E,2014ApJ...797...50A,2015ApJ...802...53A} lend considerable support to a scenario in which a significant fraction of the released energy is channelled into accelerated electrons which, guided by the surrounding magnetic field, propagate downward toward the solar surface, producing bremsstrahlung hard X-ray (HXR) emission in collisions with ambient ions along their path \cite{2011SSRv..159..301K} and heating the surrounding atmosphere through collisions with ambient electrons \cite{1988psf..book.....T}. This heating of the lower (chromospheric) layers of the solar atmosphere in turn leads to enhanced radiation in Extreme Ultraviolet (EUV) and optical wavelengths and, as a result of the associated increase in gas pressure, to an upward motion of material into the corona \cite{1989ApJ...341.1067M}.

Plasma motions (both inflows and outflows) driven by the primary magnetic reconnection process \cite{2000mare.book.....P} are also observed, both spectroscopically \cite{2013ApJ...774..122H} and through reconfiguration of the magnetic field geometry \cite{2002A&ARv..10..313P}.
The Reynolds number in the solar corona is, as in most astrophysical environments,
very large, and accordingly it is expected that these flows will be turbulent \citep{1986RvMP...58..741T,1984A&A...137...63H}.
Theoretical studies \cite{2015RSPTA.37340144L} and numerical simulations of magnetic reconnection, on both fluid \citep{2009ApJ...700...63K} and kinetic \citep{2011NatPh...7..539D,2015NatPh..11..690L} scales, have suggested that turbulence can dramatically affect the dynamics of the reconnection process. Further, magnetohydrodynamic (MHD) turbulence has long been conjectured to play a key role
in the acceleration of  particles during flares \cite{1957PhRv..107..830P,1979AIPC...56..135R,1993ApJ...418..912L,1996ApJ...467..454L}, and numerous models of turbulent (stochastic) acceleration  \citep{1997JGR...10214631M,2012SSRv..173..535P,2012ApJ...754..103B} have been proposed.

Together, the above strongly suggests a scenario in which MHD turbulence generated during magnetic reconnection plays a key role in the acceleration of particles; however, to date little firm observational evidence in support of such a scenario has been presented. In this letter, we present multi-faceted observations of an unusually well-observed solar flare that allow an evaluation of the energy content in turbulent plasma motions and hence of the role of such motions in the conversion of magnetic energy to acceleration of fast particles.

\begin{figure}[pht]
\centering
\includegraphics[trim = 10mm 0mm 0mm 15mm,clip,width=0.43 \textheight]{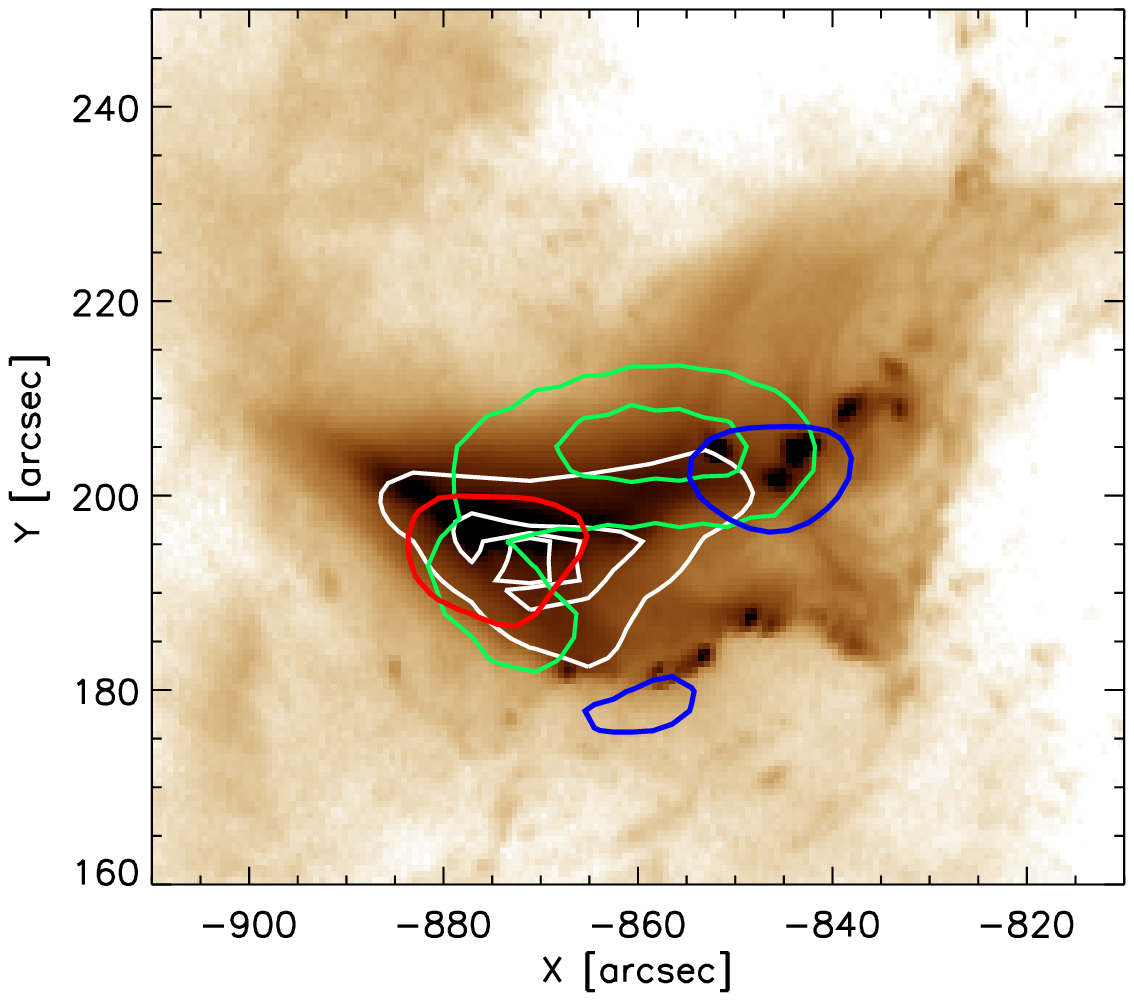}
\includegraphics[trim = 0mm 0mm 0mm 0mm,clip,width=0.3\textheight]{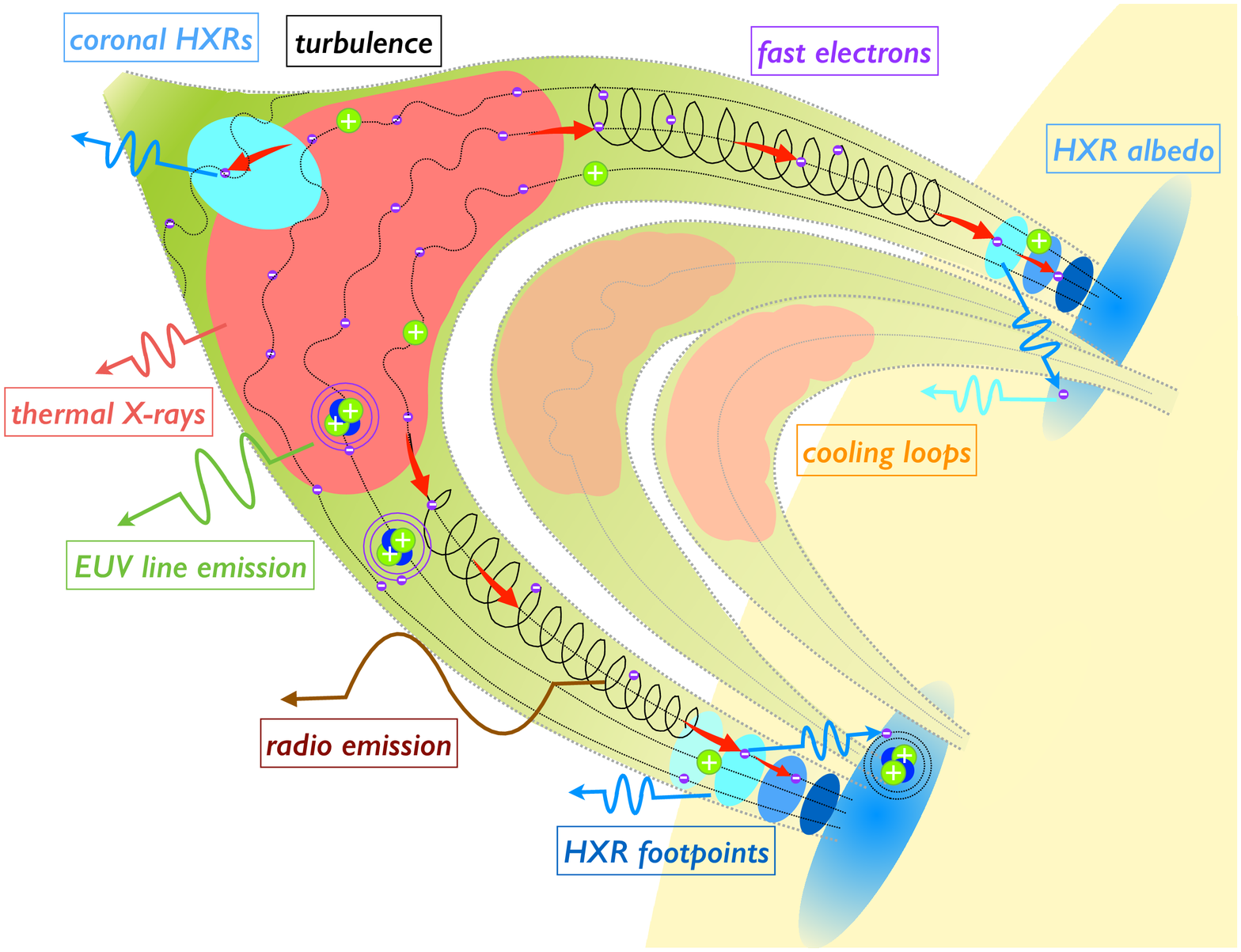}
\caption{Morphology of the flare.  \textit{Top:}
SDO/AIA 193 \AA\ image (background);
RHESSI X-ray contours at 50\% of peak value for 6 - 15 keV (red) and 25 - 50 keV (blue) energy ranges,  EIS Fe XXIV (255~\AA) intensity map (white contours at 30\% and 75\% of peak value), and Nobeyama 34~GHz radio emission (green contours at 30\% and 75\% of peak value).  \textit{Bottom:} Cartoon showing the different flare elements and the cooling post-flare magnetic loops.}
\label{fig:maps}
\end{figure}

\begin{figure}[pht]
\centering
\includegraphics[width=0.33\textheight]{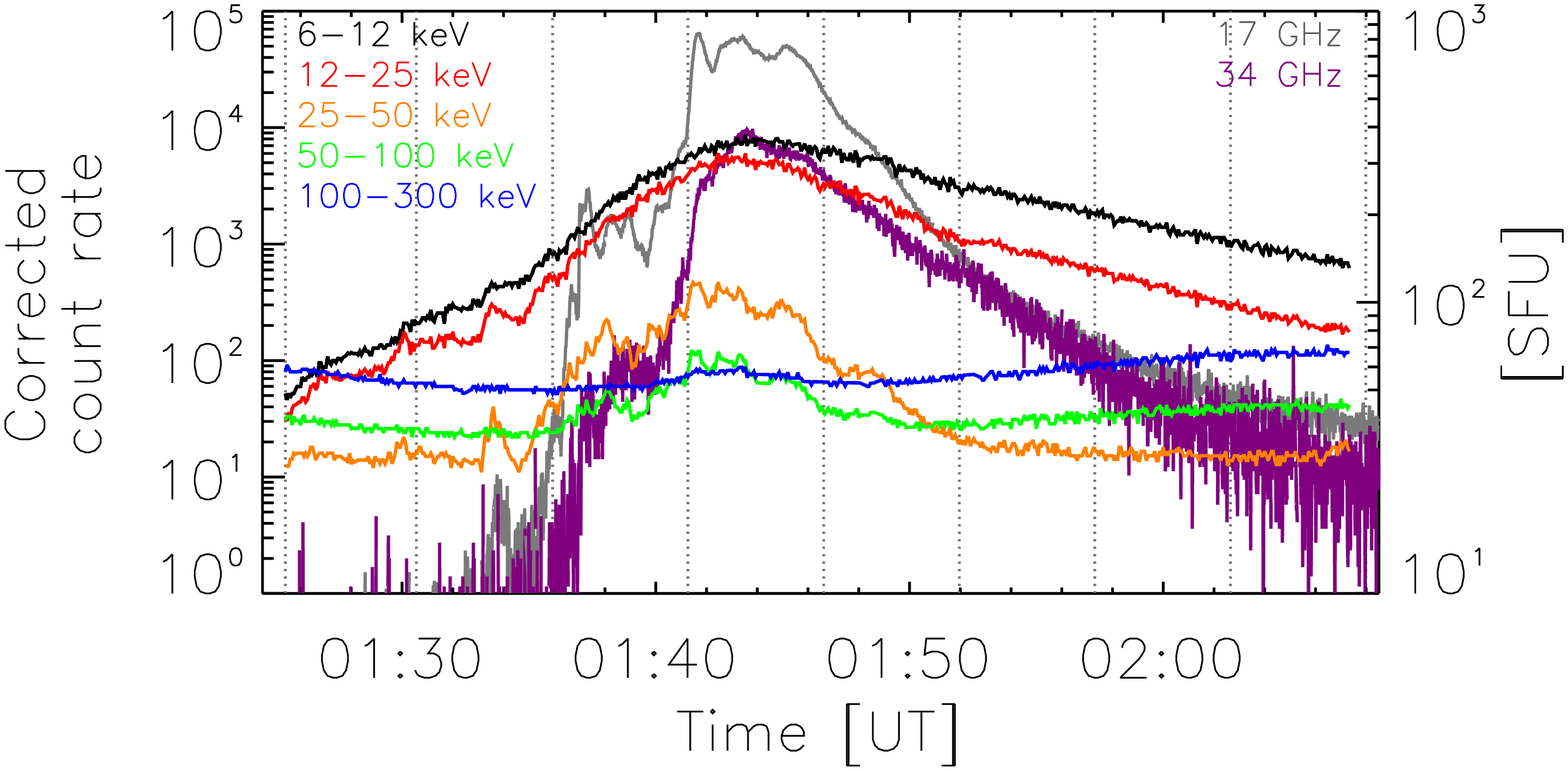}\\
\includegraphics[width=0.33\textheight]{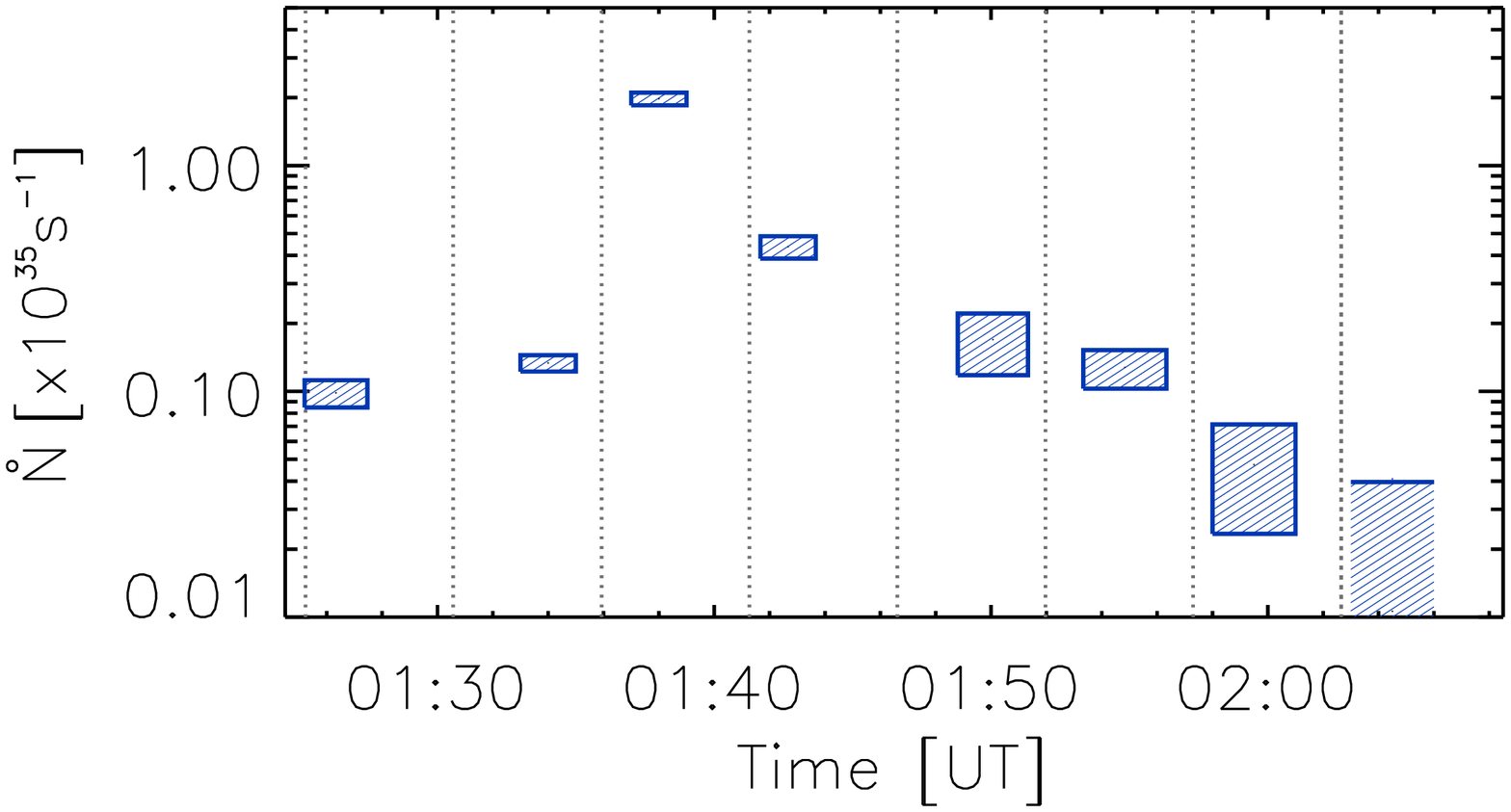}\\
\includegraphics[width=0.33\textheight]{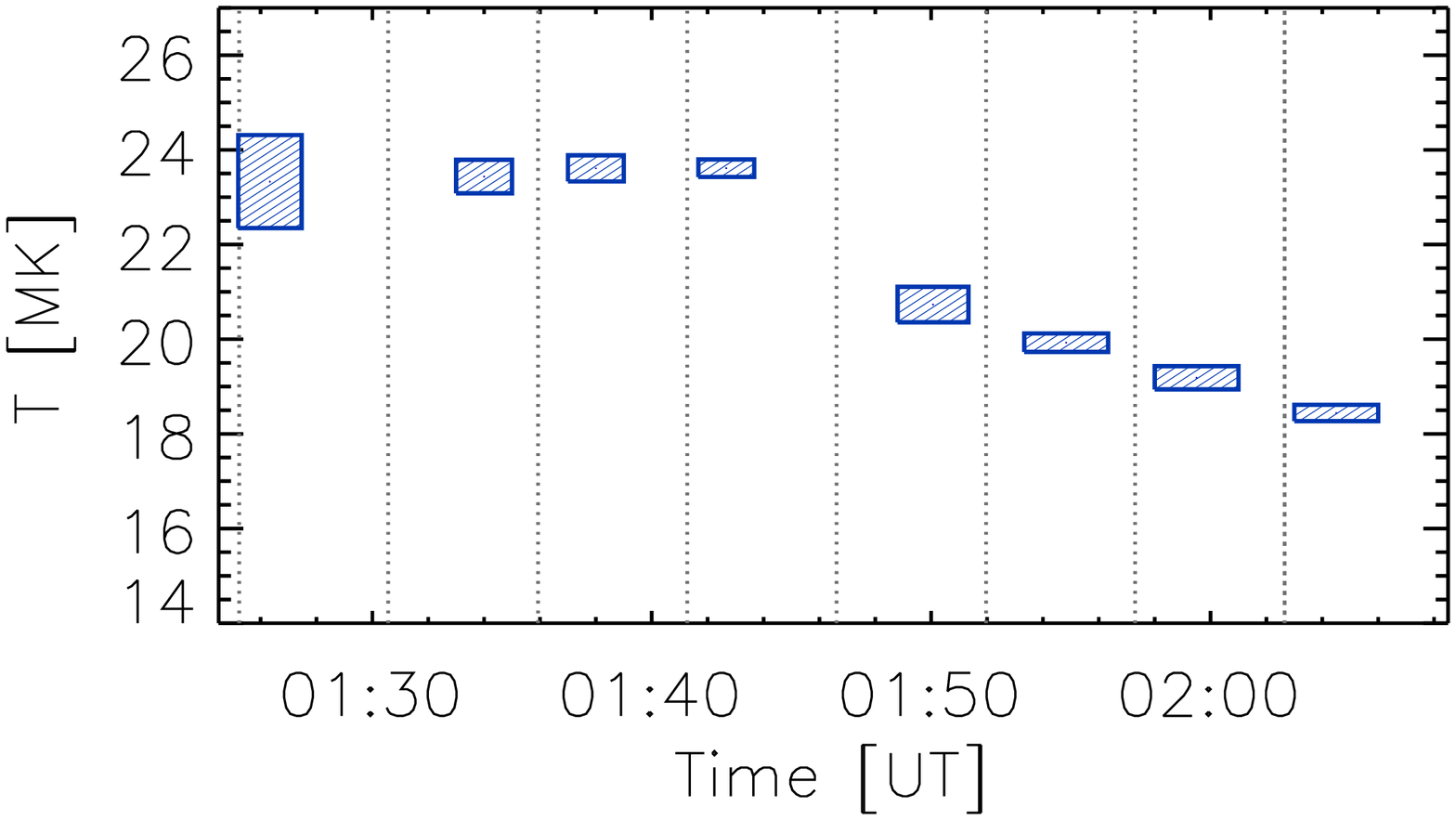}\\
\includegraphics[width=0.33\textheight]{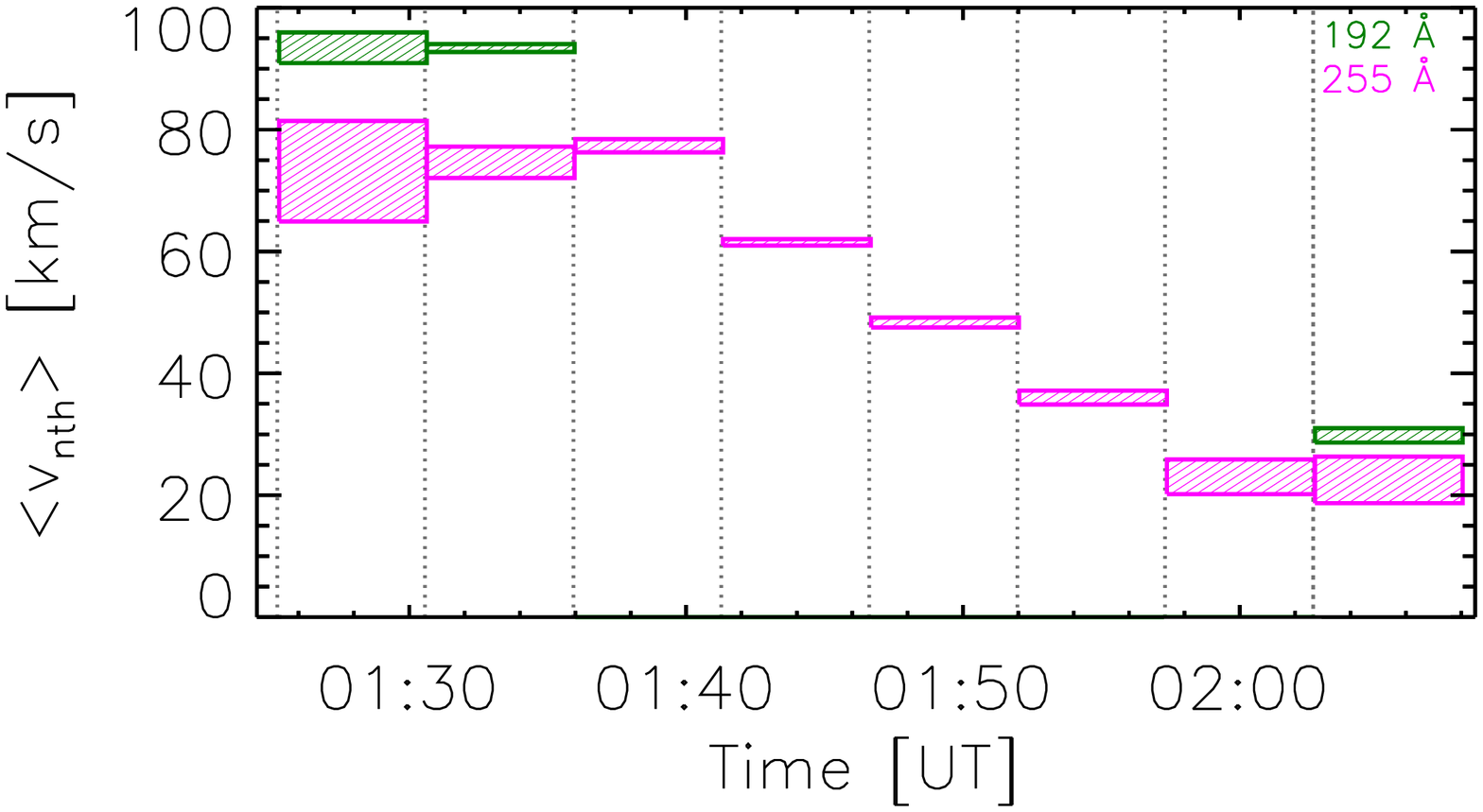}
\caption{Temporal evolution of the 2013 May 15 solar flare parameters. \textit{Top to bottom:} RHESSI X-ray and Nobeyama radio lightcurves, acceleration rate of non-thermal electrons $\dot{N}$ (from RHESSI HXR data), plasma temperature $T$ (from RHESSI SXR data), and the non-thermal broadening velocity $\langle v_{nth}\rangle$ (from Hinode/EIS) averaged over the area within the 50\% (6 - 15)~keV contour shown in Fig.~\ref{fig:maps}. The grey dotted vertical lines show the beginning and end of each EIS raster time, and the vertical range of each box indicates the uncertainty in the quantity.}
\label{fig_time-history}
\end{figure}

\begin{figure}[pht]
\includegraphics[trim = 23mm 7mm 35mm 0mm, clip, width=0.33\textheight]{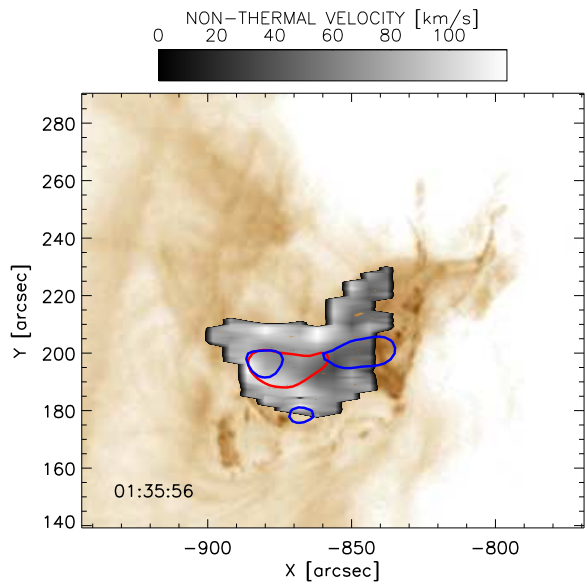}
\caption{Non-thermal velocity broadening map for the time interval 01:35:56 UT (see Fig.~\ref{fig_time-history}). Background: SDO/AIA 193 \AA. Grey scale material: EIS Fe~XXIV (255 \AA) non-thermal broadening velocity map. Red contour: 50\% of maximum intensity in $6-15$~keV HXR, blue contour: 50\% of maximum intensity at $25-50$~keV HXR.}
\label{Fig:vntmaps}
\end{figure}

\begin{figure*}[pht]
\centering
\includegraphics[width=0.31\textheight]{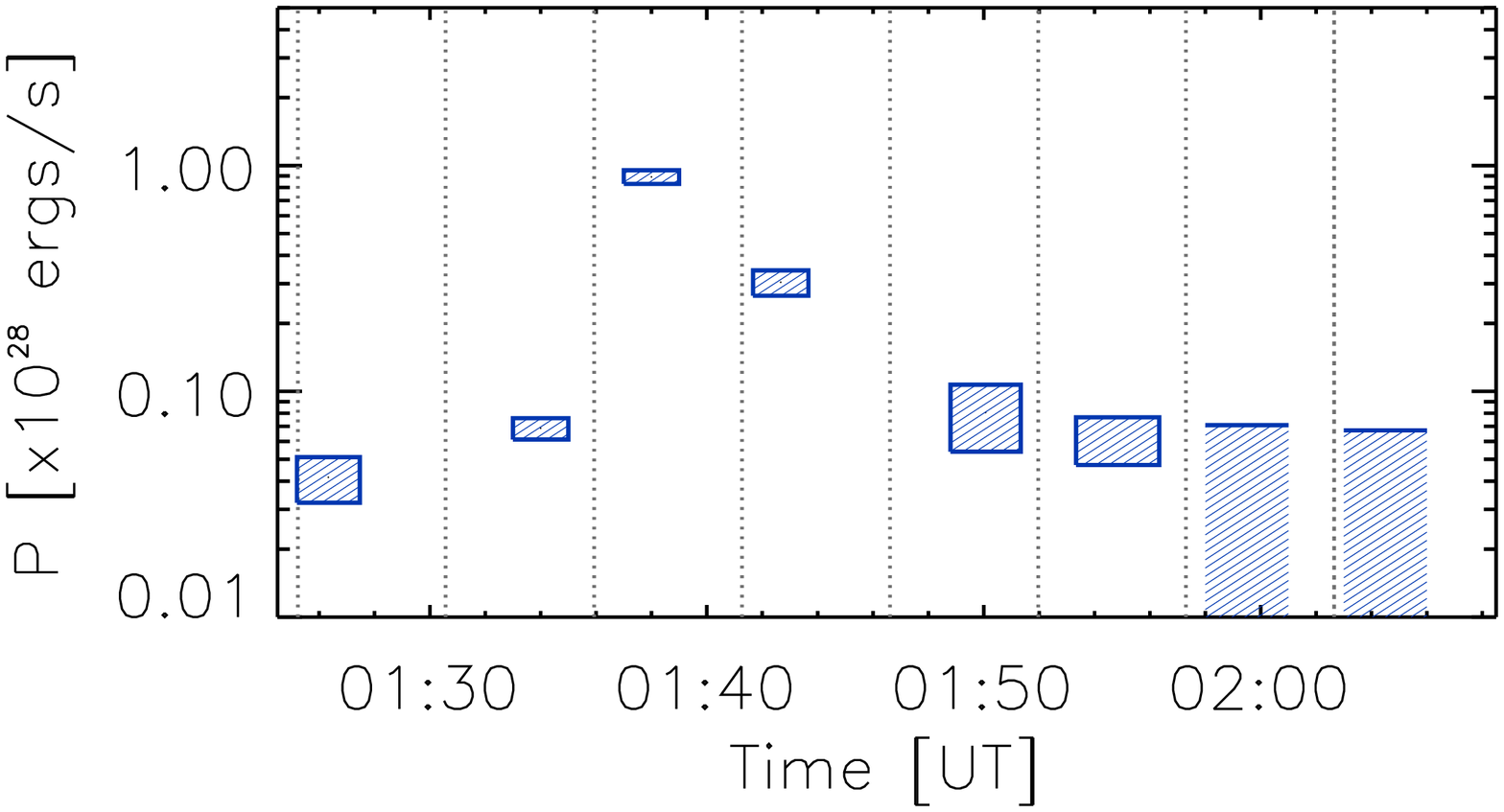}
\includegraphics[width=0.31\textheight]{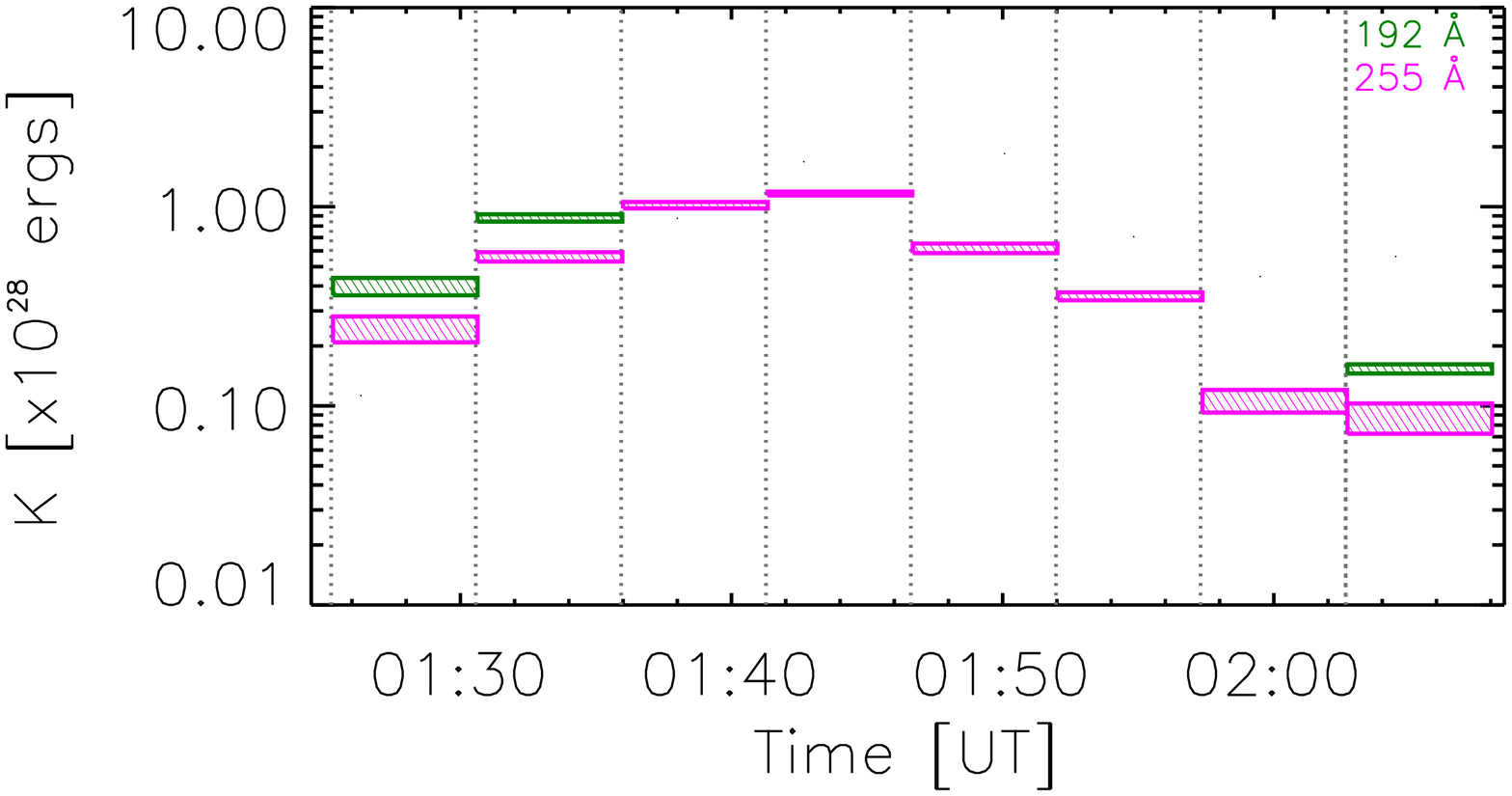}
\includegraphics[width=0.31\textheight]{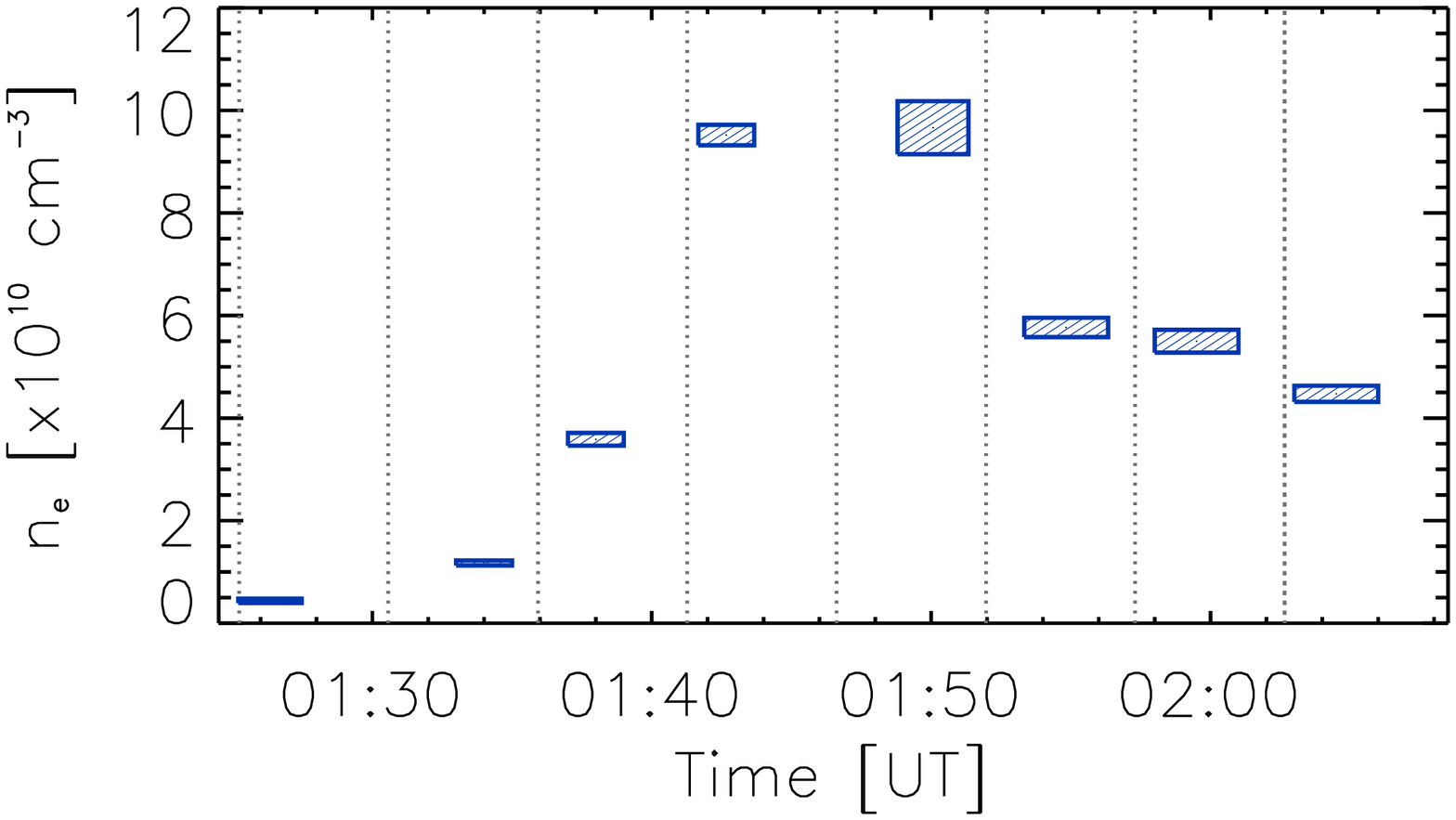}
\includegraphics[width=0.31\textheight]{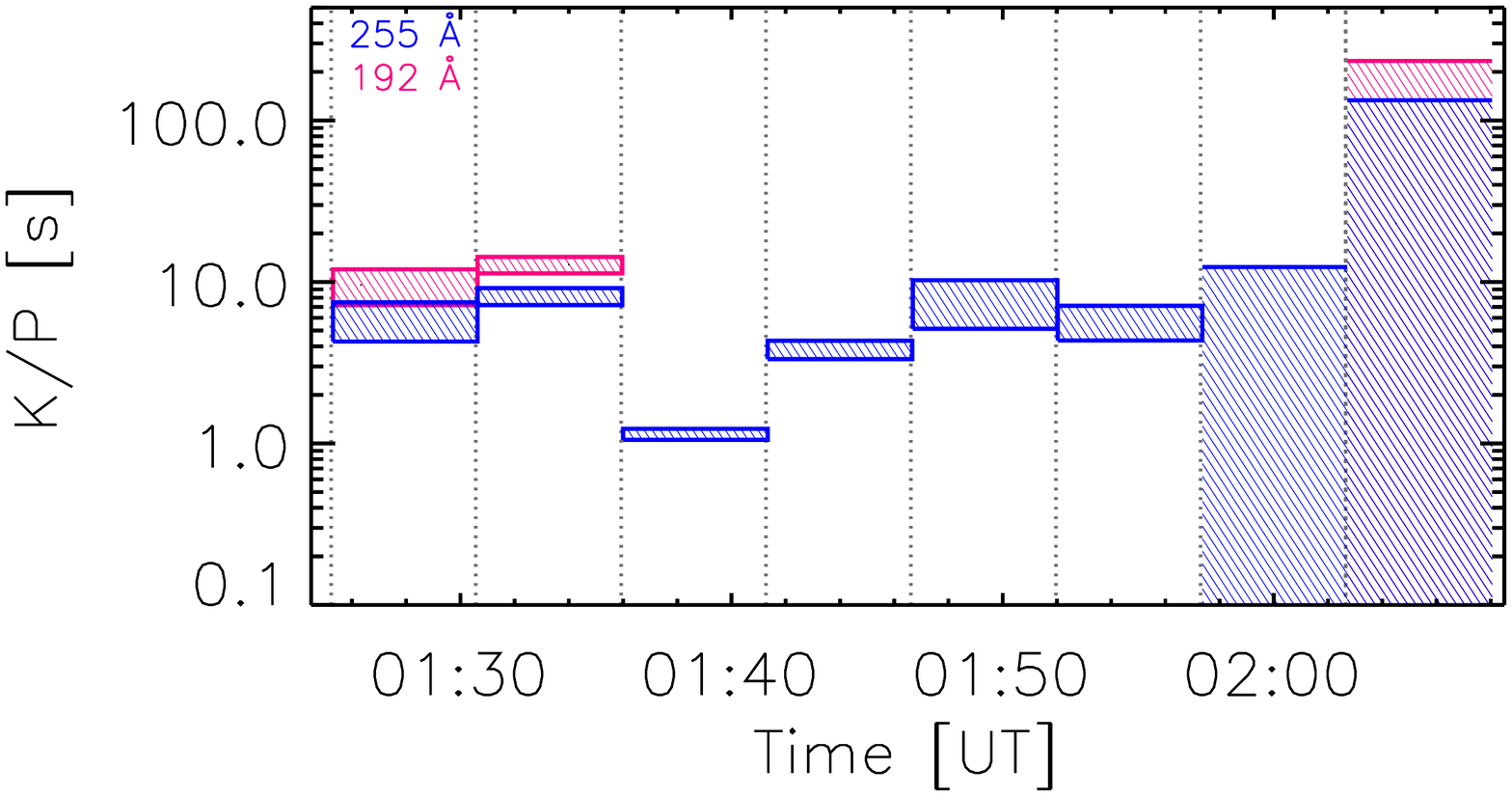}
\includegraphics[width=0.31\textheight]{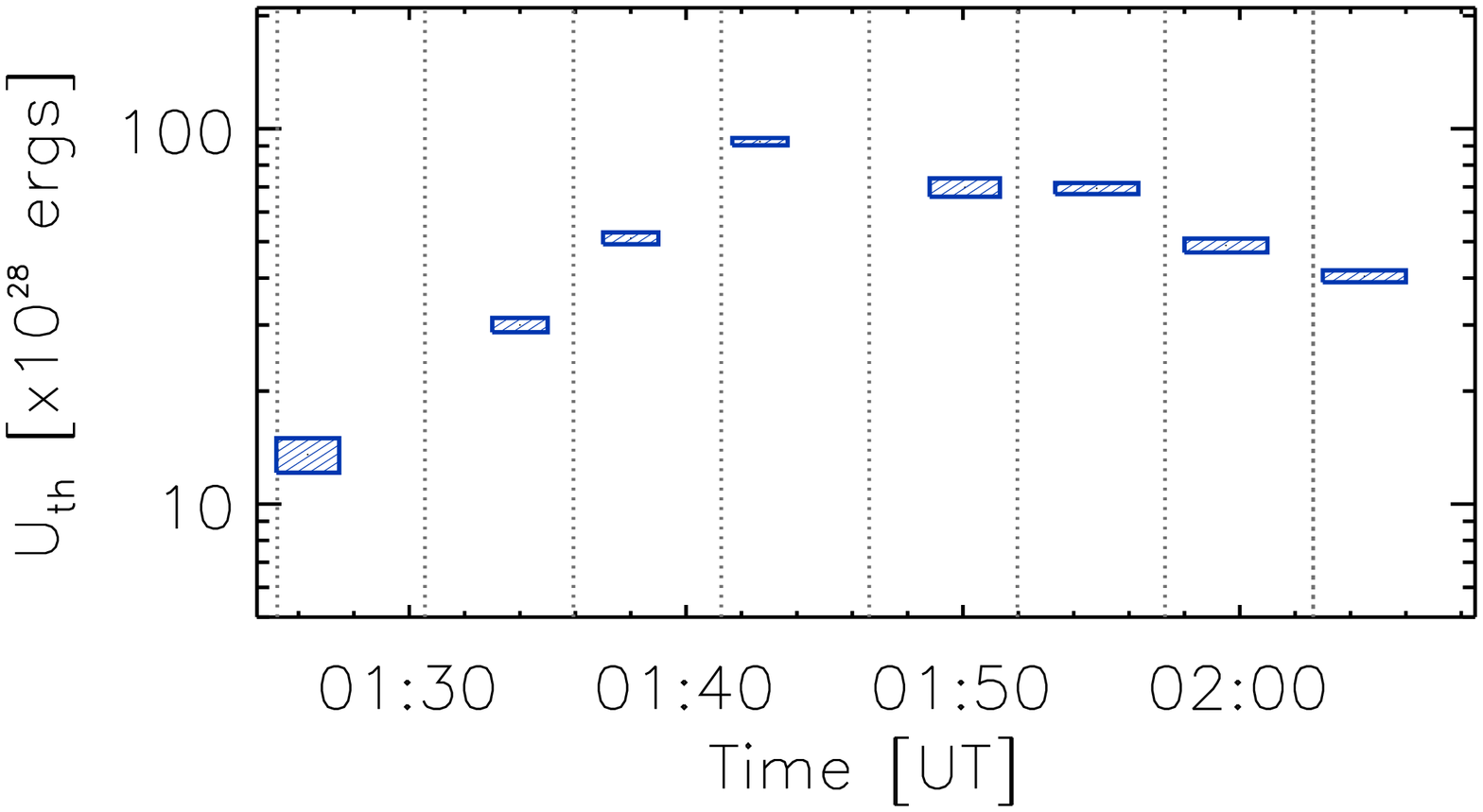}
\includegraphics[width=0.31\textheight]{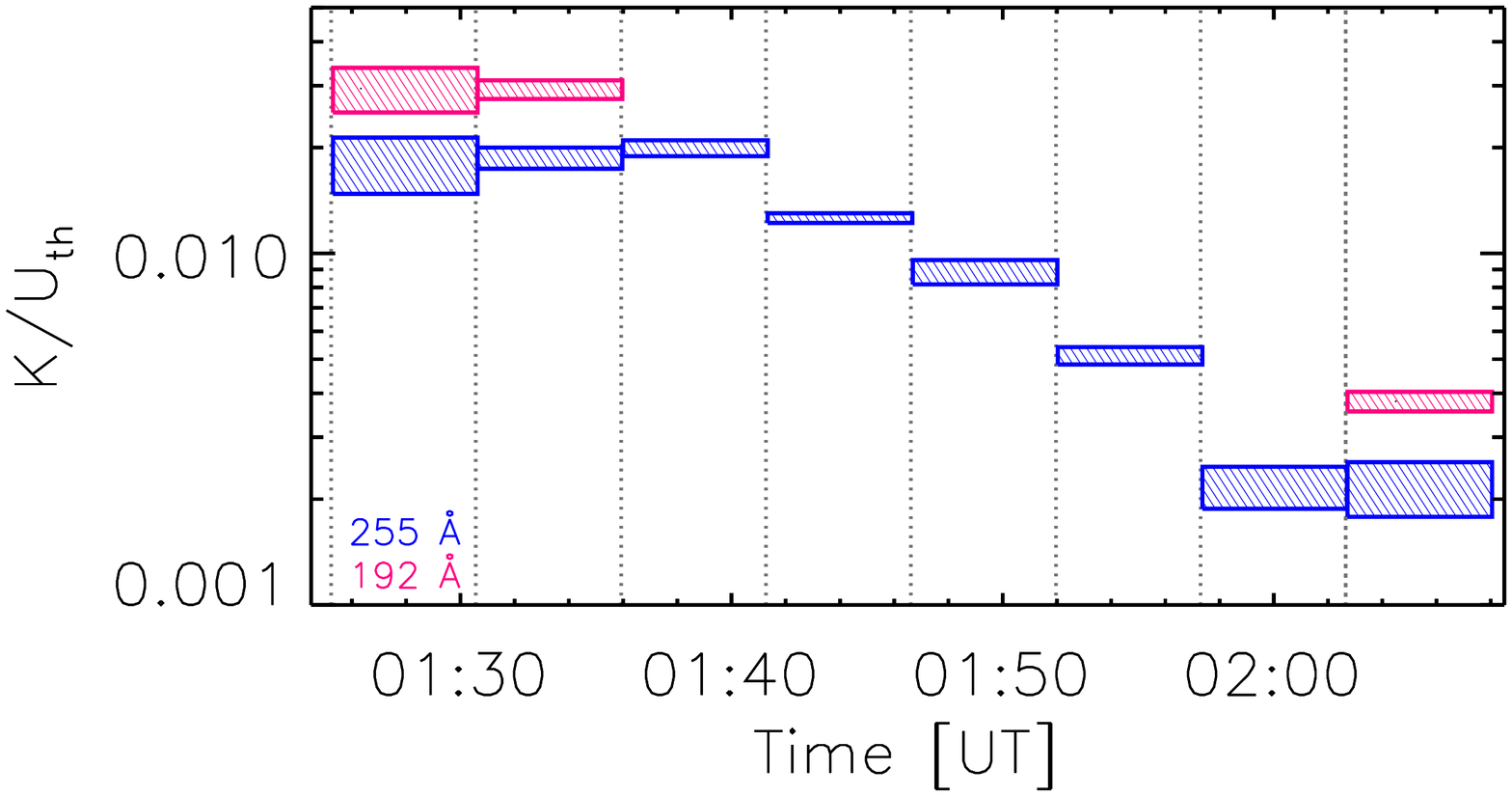}
\caption{Flare energetics.  {\it Left panels, top to bottom:} power $P$ (erg~s$^{-1}$) in non-thermal electrons above the low-energy cutoff $E_c$, density $n$ (cm$^{-3}$) of the SXR-emitting plasma, thermal energy content $U_{th}$ (erg) of the SXR-emitting plasma. {\it Right panels, top to bottom:} bulk kinetic energy $K$ (erg), ratio $K/P$ (s), ratio $K/U_{th}$ (dimensionless).}
 \label{fig:energies}
\end{figure*}

A moderately large (GOES class X1.2; see \cite{1991SoPh..133..371K}) flare occurred on 2013 May~15 in NOAA solar active region 11748. This flare was observed by several instruments: (1) the Ramaty High Energy Solar Spectroscopic Imager (RHESSI) \cite{2002SoPh..210....3L}, which produces full-Sun high-spatial-resolution soft X-ray (SXR) and hard X-ray (HXR) images with $\sim$1~keV spectral resolution and several-second time resolution; (2) the Solar Dynamics Observatory (SDO) Helioseismic and Magnetic Imager (HMI) \cite{2012SoPh..275..207S}, which measures the magnetic field in the lower atmospheric levels; (3) the SDO Atmospheric Imaging Assembly (AIA) \cite{2012SoPh..275...17L}, which provides high-resolution EUV spatial images; (4) the Hinode EUV Imaging Spectrometer (EIS) \citep{2007SoPh..243...19C}, which produces EUV spectral line profiles; and (5) the Nobeyama Radioheliograph and Radiopolarimeters \cite{1994IEEEP..82..705N}, which measure the radio-wave radiation produced by mildly relativistic electrons and so provides a diagnosis of the magnetic field strength in the corona.

Fig.~\ref{fig:maps} shows the general morphology of the flare consistent with the flare reconnection geometry\cite{2002A&ARv..10..313P}; its near-limb location allows us to readily ascertain its vertical structure. The flare has a cusp-shaped coronal structure, clearly visible in the AIA 193~\AA\ image which delineates hot flare plasma with temperature $\sim 10^{7.3}$~K; the EIS Fe~XXIV~192~\AA\ and~255~\AA\ line intensities, which both delineate plasma with temperature $\sim 10^{7.2}$~K, both exhibit a similar structure. RHESSI observations reveal a bright coronal (loop-top) SXR ($6-15$~keV) source and two HXR ($\gtrsim$25~keV) footpoints where the accelerated electrons, traveling along the magnetic field lines, impact the relatively dense chromospheric layers of the atmosphere. Using the EUV and SXR images, we estimate the height of the magnetic loop to be $\sim$$1.5 \times 10^4$~km.

Fig.~\ref{fig_time-history} shows the temporal evolution of the flare in radio, SXRs and HXRs. The SXR emission has a spectral shape consistent with bremsstrahlung from a Maxwellian distribution of electrons in a thermal plasma with $T \sim 10^{7.5}$~K, while the HXR emission is produced by bremsstrahlung from higher energy ($\gtrsim$25~keV) non-thermal electrons and typically has a harder, power-law, spectral shape.  The peak in HXR emission occurred at $\sim$01:41~UT, followed at $\sim$01:45~UT by the peak in SXR emission. The main microwave peaks at 17 and 34~GHz were observed by Nobeyama at $\sim$01:41~UT, near the time of the peak in HXRs.

Following the methods described in \citep{2011SSRv..159..107H,2011SSRv..159..301K}, which include consideration of the primary bremsstrahlung emission mechanism, instrumental pulse pile-up effects, and the albedo flux resulting from photospheric reflection of primary source HXRs, HXR observations allow us to infer the spatial and spectral distributions of the emitting energetic electrons (Fig.~\ref{fig_time-history}). The rate of production,
$\dot{N}$, of accelerated electrons above a specified energy $E$ is roughly proportional to the overall intensity of the HXR spectrum above that energy.
Because the HXR spectrum, and hence the accelerated electron spectrum that produces it, is quite steep ($\propto E^{-\delta}$ with a power-law index $\delta$
typically $\sim$4-6), the total energy in accelerated electrons depends
on the shape of the low-energy end of the HXR spectrum.
We selected time intervals to avoid instrumental effects such as RHESSI shutter motions  and applied the methodology in \cite{2015ApJ...809...35K} to produce a range of values of $\dot{N}$ consistent with data at each time interval throughout the flare, as shown by the vertical extent of the boxes in Fig.~\ref{fig_time-history}.

The emission measure EM=$n^2 V$ and temperature $T$ of the hot thermal SXR-emitting plasma are determined from an isothermal fit \cite{2011SSRv..159..107H} to the SXR spectral component; their variations throughout the event are shown in Fig.~\ref{fig_time-history}.
Using the inferred value of EM and the source volume $V$ estimated from
Fig.~\ref{fig:maps} leads to a estimate of the source density $n$,
which is the lower limit because of the possibility that the emission originates only from a fraction of the observed flare volume, although the estimates \cite{1990ApJS...73..117D} indicate that this ``filling factor'' is consistent with unity.

Broadening of spectral lines in excess of the thermal Doppler width \cite{1987ApJ...313..449B} is a signature of turbulent motions,
associated either with plasma oscillations \cite{2012SSRv..173..535P}
or unresolved bulk plasma motions \cite{1989ApJ...341.1067M}.
To estimate the extent of such turbulence, we use the EIS Fe~XXIV spectral line profiles at 192~\AA\ and 255~\AA. Since the Fe~XXIV 192~\AA\ line represents some 80\% of the total intensity in the AIA 193~\AA\ passband,
the EIS and AIA images were co-aligned by cross-correlating the EIS 192~\AA\ intensity maps with the AIA 193~\AA\ images.
Then, the EIS Fe~XXIV 255~\AA\ line profile at each point in the image was fitted using a Gaussian, following the procedure in \cite{2013ApJ...774..122H},
which allows for instrumental effects. The characteristic non-thermal broadening velocity $v_{nth}$ is then determined from the extent to which the observed spectral line width exceeds that expected from thermal line broadening \citep{2013ApJ...774..122H}. Fig.~\ref{Fig:vntmaps} shows the spatial distribution of the 255~\AA\ non-thermal line-broadening velocities throughout the source for the time interval starting $01:35:56$ UT (Fig.~\ref{fig_time-history}), which corresponds to the interesting epoch just prior to the peak in the HXR light curve. Pixels were excluded where the intensity was either too strong (saturated) or too weak for $v_{nth}$ to be reliably determined.  The turbulent velocity tends to be larger by $\sim$50\% near the apex of the magnetic loops and along the outer edge of the arcade \cite{2014ApJ...788...26D}. Fig.~\ref{fig_time-history} shows the time variation of $\langle v_{nth}\rangle$, the value averaged over the area $A$ inside the 50\% contour of the RHESSI HXR (6 - 15~keV) map (Fig.~\ref{Fig:vntmaps}).

A similar procedure was used for the Fe XXIV 192~\AA\ line; however, this line was more strongly saturated \cite{2014ApJ...788...26D} and hence useful measurements were available only near the start and the end of the flare. Where information from both Fe XXIV 255~\AA\ and 192~\AA\ lines were available (i.e., before 01:36 UT and after 02:03 UT), the inferred values of the average non-thermal broadening velocity $\langle v_{nth}\rangle$ agreed within $10\%$ (Fig.~\ref{fig_time-history}). Typical values of $\langle v_{nth}\rangle$ in this $10^{7.2}$~K plasma were found to be $(60-100)$~km~s$^{-1}$. This is somewhat lower than the previously reported (spatially-unresolved) measurements of $\langle v_{nth}\rangle \simeq 200$~km~s$^{-1}$ at higher temperatures \cite{1982SoPh...78..107A,1987ApJ...313..449B,1990ApJS...73..117D,1995ApJ...438..480A},
suggesting, not surprisingly, that hotter plasma may admit higher turbulent velocities. The total turbulent kinetic energy $K\propto \langle v_{nth}\rangle ^2$ could therefore be larger by a factor of $\sim$4 than that inferred from the $10^{7.2}$~K lines alone.

The power in non-thermal electrons is given by $P=((\delta -1)/(\delta -2)) \dot{N} ̇E_c \,$; its time history closely matches that of the HXR flux. RHESSI images (Fig.~\ref{fig:maps}) show both the location and the area $A$ of the coronal source, deduced from the 50\% intensity contour in the (6-15)~keV map, allowing an estimate of the source volume $V = A^{3/2} \simeq 2\times 10^{27}$~cm$^{3}$, and hence \cite{2012ApJ...759...71E} the thermal plasma energy in the coronal source $U_{th} = 3kT\sqrt{{\rm EM} \cdot V}$, where $k$ is Boltzmann's constant.

The turbulent kinetic energy $K$ at each EIS raster time is calculated using $K = (3/2) m_i \, \langle v_{nth}\rangle^2 \, n_p \, V $, where $m_i = 1.3 \, m_p$ is the mean ion mass for solar coronal abundances \cite{2014SoPh..289..977R} and $n=\sqrt{EM/V}$ is the number density. Fig.~\ref{fig:energies} shows the ratio of the turbulent kinetic energy $K$ (ergs) to the instantaneous thermal energy content $U_{th}$ (ergs); the temporal behavior of $K$ is similar to that of $U_{th}$, with $K$ some two to three orders of magnitude smaller, varying between $10^{27}$ and $10^{28}$~ergs. Since both $K$ and $U_{th} \propto nV \sim \sqrt{EM \cdot V}$, the effect of a volumetric filling factor less than unity is to reduce them both somewhat, but the ratio $K/U_{th}$ is preserved. While $K$ attains its peak value around the same time ($\sim$01:25~UT) as the SXR flux, it notably has a value equal to some 20\% of its peak value as early as 01:40~UT, well before the peak in the HXR flux. Similar behavior is also seen in MHD simulations \cite{2016A&A...589A.104G}. Fig.~\ref{fig:energies} also shows the ratio of the turbulent kinetic energy $K$ (ergs) to the power $P$ (erg~s$^{-1}$) in energetic electrons; the ratio $K/P$ (which is a measure of the time it takes a power $P$ to energize/deplete a reservoir of energy $K$)  has a relatively steady value of order 1-10~s.

To estimate the available energy in the magnetic field, we used two independent techniques: (i)~microwave spectral data from NoRH and NoRP, and~(ii)~extrapolated HMI line-of-sight photospheric magnetograms.
The microwave spectra were fitted assuming isotropic electrons with a power-law energy spectrum as determined from the RHESSI HXR spectrum.
Using fast gyrosynchrotron codes \cite{2010ApJ...721.1127F}, we reproduced the observed NoRP microwave fluxes at three frequencies: 17~GHz and 34~GHz (in the optically thin range) and 9.4~GHz (near the spectral peak); the best-fit spectra corresponded to an average magnetic field strength $B\simeq (300-400)$~G.  The coronal magnetic field strength was also estimated from potential-field (minimum magnetic field strength) extrapolation of the observed HMI line-of-sight photospheric magnetograms, giving $B\simeq 300$~G at heights $\sim 1.5\times 10^4$~km,
where the bulk of the radio emission is observed.  These mutually consistent values of the magnetic field strength $B$ correspond to a total magnetic energy $(B^2/8\pi) \, V \simeq (7-12)\times 10^{30}$ erg.
Following \cite{2012ApJ...759...71E}, we estimate that the magnetic energy available for dissipation (i.e., the excess over the potential field energy) is 30\% of the total magnetic field energy, or $\sim 2 \times 10^{30}$~erg.

An enduring challenge in flare physics relates to how such a large fraction of the stored magnetic energy is converted to energy in accelerated particles.
In relation to particle acceleration, there are, broadly-speaking,
two representations of turbulence (stochasticity): ``wave turbulence'' \cite{1968Ap&SS...2..171M,1987SoPh..113..195M,1993ApJ...418..912L}
and a ``stochastic-ensemble-of-current-sheets'' \cite{2004ApJ...605L..69A,2014A&A...561A..72G,2016ApJ...827L...3V,2016A&A...589A.104G}. These two concepts are not necessarily unrelated, since there is a tendency for MHD turbulence to form current sheets \cite{2004ApJ...617..667D}.
In both scenarios, energy produced at large scales systematically cascades to smaller and smaller scales, where the energy is eventually dissipated to produce heating and acceleration of non-thermal particles.
The rate of energy release at large scales and the rate of subsequent energy transfer to smaller, dissipative, scales together determine the rate at which particle acceleration can occur.

In light of this discussion, two aspects of the turbulent energy content $K$ inferred herein are significant.  First, the turbulent energy is observed (Figs.~\ref{fig:maps} and~\ref{Fig:vntmaps}) to be spatially concentrated in the coronal part of the magnetic loop below the observed cusp-like structure, where the primary energy release is believed to occur.  Second, its energy content $K$ grows to a significant level well before the peak in HXR intensity, i.e., before the maximum rate of electron acceleration (Fig.~\ref{fig:energies}). Together, these features lead us to propose that turbulence constitutes a viable channel for the conduit of cascading energy. Although the instantaneous turbulent energy content $K$ is only a percent or so of the available magnetic energy (and of the thermal energy $U_{th}$ in the SXR-emitting plasma), the transfer of energy out of the turbulent energy reservoir
could be sufficiently rapid for the associated power to rival that associated with dissipation of the turbulence and the acceleration of non-thermal particles.
The ratio of $K/P$ (Fig.~\ref{fig:energies}) shows that for such a scenario
to be viable the turbulent energy must be dissipated (and replenished) on a timescale $\sim$1-10~s. Such a timescale is consistent not only with observed fluctuations in the time profile of the HXR emission in the event studied here, but also with many previous studies \cite{1988psf..book.....T,2011SSRv..159..107H}.

It is well known \cite{1995ApJ...438..763G} that dissipation of anisotropic Alfv\'en MHD turbulence occurs on a timescale $L_\perp/\langle v_{nth} \rangle$, where $L_\perp$ is the characteristic scale associated with variations $\delta B$ perpendicular to the guiding magnetic field.
The ``side-on'' geometry of this particular flare on the sky (Fig.~\ref{fig:maps}) suggests that the observed line-of-sight velocity fluctuations $\langle v_{nth}\rangle$ correspond to motions perpendicular to the guiding field.
And although $L_\perp$ is not directly observable, the dissipation timescale can nevertheless be estimated as follows \cite{1995ApJ...438..763G}.
The energy density associated with a turbulence-perturbed magnetic field $\delta B$ is $U_B \simeq (\delta B)^2/8 \pi$. Equating this to the turbulent energy content $K = (1/2) n \, m \, \langle v_{nth}^2\rangle$, we obtain $\langle v_{nth}^2\rangle \simeq (\delta B)^2/4 \pi n \, m$.  Since the Alfv\'en speed $V_A = \sqrt{B^2/4 \pi n \, m}$, it follows that $\langle v_{nth}\rangle/V_A\simeq \delta B/B \simeq L_\perp/L_\parallel$, where $L_\parallel$ is the longitudinal extent of the turbulence region. Thus the dissipation timescale $L_\perp/\langle v_{nth} \rangle$ is approximately the same as the Alfv\'en crossing time $L_\parallel/V_A$,
a quantity that is readily ascertainable from observations.
Using the inferred values of $B$ and $n$ gives $V_A \simeq 2\times 10^3$~km~s$^{-1}$ for this flare, a typical value for the flaring corona \cite{2011SSRv..159..107H}.
Thus we expect dissipation of turbulent energy to occur on a timescale
$L_\parallel/V_A \simeq 5$~s, a value consistent both with the inferred value
of $K/P$ (Fig.~\ref{fig:energies}) and with the timescales typically associated with the acceleration of electrons by MHD wave turbulence \citep{1996ApJ...461..445M}.
In the stochastic current sheet models, the ratio $U_B/K$ could be different from one, but the observed $K$ could be used to test these models.

In summary, the suite of observations presented herein demonstrate the presence, in the acceleration region, of a significant energy reservoir
in turbulent plasma motions which correlates well in time with the acceleration of HXR-producing electrons. An instantaneous energy content $\sim$$10^{28}$ ergs, produced and dissipated on a timescale of a few seconds,
transfers a steady-state power $\sim(0.1-1)\times10^{28}$~erg~s$^{-1}$, rivalling the power in accelerated non-thermal particles.  These observations not only enable quantitative testing of turbulence acceleration models;
they lend considerable credence to the idea that turbulence acts as a crucial intermediary in the transfer of energy from reconnecting magnetic fields
to accelerated particles during solar flares, and therefore presumably in other astrophysical particle acceleration sites.

\begin{acknowledgments}
The authors are thankful to the anonymous two referees for constructive comments and criticism and S. Sridhar for useful discussions. Hinode is a Japanese mission developed and launched by ISAS/JAXA, with NAOJ as a domestic partner and NASA and UKSA as international partners. It is operated by these agencies in cooperation with ESA and NSC (Norway). EPK, NLSJ, NHB were supported by a STFC consolidated grant ST/L000741/1. JEP was supported by CONACyT postdoctoral grant 207887. AGE was supported by grant NNX10AT78G from NASA's Goddard Space Flight Center. AAK was supported in part by the RFBR grant 15-02-03717.

\end{acknowledgments}

\bibliography{vnt_references}

\end{document}